\documentclass[twocolumn,trackchanges]{aastex701}
\usepackage{amsmath}

\begin{document}

\title{Exploring the Model Dependence of MCMC-Based 21 cm Power Spectrum Parameter Constraints}

\author[orcid=0009-0005-4961-3013]{A. Berklas}
\affiliation{Department of Physics, Brown University, Providence, 02912, Rhode Island, United States}
\email[show]{august\_berklas@brown.edu}  

\author[orcid=0000-0002-3492-0433]{J. C. Pober} 
\affiliation{Department of Physics, Brown University, Providence, 02912, Rhode Island, United States}
\email{jonathan\_pober@brown.edu}

\begin{abstract}

Detection and analysis of the cosmic 21 cm signal of neutral hydrogen has long been considered the most promising route towards exploration of the Epoch of Reionization (EoR). 21CMMC, a Markov Chain Monte Carlo sampler of the seminumerical simulation code 21cmFAST, has already been used in conjunction with published upper limits on the 21 cm signal from the Murchison Widefield Array, the Low Frequency Array, and the Hydrogen Epoch of Reionization Array to constrain the astrophysics of the EoR. Here, we investigate the extent to which analysis of the EoR performed using 21CMMC is dependent on the underlying seminumerical model. We used 21cmFAST to simulate two datasets of 21 cm light-cones that differ only in the algorithm used to identify ionized regions (the so-called “bubble-finding” algorithm). We then tested 21CMMC’s ability to return key astrophysical parameters when using the different bubble-finding algorithms. We find that the performance of 21CMMC depends sensitively on the agreement between the astrophysical model of our mock data and the model used for sampling. This result has important implications for the analysis of the 21 cm signal performed using 21CMMC and further motivates investigation into model-independent analysis techniques for 21 cm EoR data.

\end{abstract}


\section{Introduction} 
\label{sec:introduction}

The Epoch of Reionization (EoR) describes the period in cosmic history during which neutral hydrogen, which filled the Intergalactic Medium (IGM), was ionized by some of the earliest luminous sources. Although the first stars in the Universe formed during the preceding cosmic dawn, it is only during the EoR that enough ultraviolet radiation was emitted to enable the IGM to undergo the major baryonic phase change that characterizes its present state. Analysis of the EoR, therefore, provides insight into the evolution of large-scale structure in the early Universe.

One of the most promising avenues towards studying the EoR is through detection of the hyperfine transition of neutral hydrogen and the measurement of its power spectra (PS). During the EoR, luminous sources inhomogeneously ionize the IGM, creating ionized bubbles which expand as reionization progresses. This spatial inhomogeneity leaves a detectable imprint on the 21 cm PS. Reviews of ``21 cm cosmology'' and the associated techniques can be found in \citet{furlanetto_et_al_2006}, \citet{morales_and_wyithe_2011}, and \citet{liu_and_shaw_2020}.

Unfortunately, detection of the 21 cm signal has proven difficult due to the weakness of the signal in comparison to the astrophysical foregrounds, human-generated interference, and instrumental noise. As of yet, there has only been one unconfirmed detection of the global 21 cm signal, which was made by the Experiment to Detect the Global EoR Signature (EDGES; \citet{EdgesA}). There have been no detections of the 21 cm PS from the EoR to date. 

In spite of these challenges, published upper limits on the 21 cm PS have already been used to constrain the astrophysics of the EoR and disfavor cold models of reionization. Data from the Low Frequency ARray (LOFAR) at $z \approx 9.1$ was used to constrain spin temperature (\citet{GreigA}), although they were unable to rule out models that had not been previously disfavored by existing probes. However, data from both the Murchison Widefield Array (MWA; \citet{GreigB}) and the Hydrogen Epoch of Reionization Array (HERA; \citet{HeraA} and \citet{HeraB}) at redshifts of $z = 6.5-8.7$ and $z = 7.9 \text{ and } 10.4$ respectively, were used to weakly constrain astrophysical parameters of the EoR beyond existing constraints. 

The above-referenced studies made use of 21CMMC, a Markov Chain Monte Carlo sampler of the seminumerical 21 cm signal simulation tool 21cmFAST, which uses the power spectrum of the 21 cm signal as a summary statistic for comparison. Recent years have seen the development of image-based reionization analyses using techniques from computer vision and AI (e.g. \citealt{Gillet2019,LaPlante2019,Kwon2020,Neutsch2022}), but 21CMMC remains the default analysis for the field, both because of the heritage of MCMC analyses using the power spectrum from the Cosmic Microwave Background community, and because current and upcoming experiments lack the sensitivity for imaging and focus on a power spectrum detection. However, using 21CMMC to constrain astrophysical parameters of the EoR takes for granted that 21cmFAST is capable of capturing the underlying astrophysics of the EoR. Although studies have shown that 21cmFAST realizations of reionization agree well with more sophisticated radiative transfer simulations \citep{zahn11, Mesinger11}, there have been no demonstrations where an MCMC analysis based on 21cmFAST realizations was able to robustly interpret other reionization simulations. In particular, 21cmFAST simulations may require nonphysical values for specific parameters to reproduce the true 21 cm PS. Constraints on the astrophysical parameterization of the EoR made by 21CMMC, then, would not be accurate.

We investigate one aspect of this issue by making use of a built-in feature of 21cmFAST. 21cmFAST allows the user to select from one of two bubble-finding algorithms in simulating realizations of the 21 cm signal.\footnote{As of the release of 21cmFAST version 4.0.0 in April, 2025, both bubble-finding algorithms have been deprecated in favor of a ``discrete halo sampler'' \citet{v4}. At time of writing, this version is not compatible with 21CMMC but remains an interesting point of investigation for future work.} Features of these models are described in \citet{Mesinger07}, \citet{Zahn07}, \citet{zahn11}, and \citet{Mesinger11}, and are referred to as bubble-finding algorithms 1 and 2 going forward. The two bubble-finding algorithms differ in their treatment of ionized bubbles in the IGM as filter scale decreases with bubble-finding algorithm 1 being more computationally expensive. We describe the difference between the two bubble-finding algorithms in more detail in Section \ref{sec:algorithm_differences}. We simulate light-cones\footnote{Only loosely related to the concept from special relativity, 3D volumes of the 21 cm signal's position on the sky with the depth axis corresponding to the redshift at which it was emitted are referred to as ``light-cones.''} of the 21 cm signal using each bubble-finding algorithm, then use 21CMMC, sampling bubble-finding algorithm 1, to return values of the reionization parameters $\zeta$ and $T_{vir}^{min}$ (described in \ref{sec:21cmFAST_setup}) for light-cones simulated using bubble-finding algorithm 2 and vice versa. We also explore whether 21CMMC, sampling bubble-finding algorithm 1, can return the in-principle model-independent values of the duration and midpoint of reionization for a bubble-finding algorithm 2 light-cone. These analyses give us insight into the degree to which 21CMMC's constraining power is model-dependent, giving us intuition as to 21CMMC's constraining power on an actual 21 cm PS.

The remainder of this paper is organized as follows. In Section \ref{sec:methodology}, we describe our 21cmFAST and 21CMMC setups. We present our main result in Section \ref{sec:results}, and three follow up tests in Sections \ref{sec:PS_recreate}, \ref{sec:observables}, and \ref{sec:instrumental_noise}. In Section \ref{sec:discussion}, we present our discussion, and finally, in Section \ref{sec:conclusion}, we provide our conclusion. Unless otherwise stated, we adopt cosmological parameters in agreement with the Planck 2018 results \citep{Planck}.

\section{Methodology}
\label{sec:methodology}

We simulate the 21 cm signal using the seminumerical simulation tool 21cmFAST version 3.3.2.dev132+gaefc273.d20240617 (\citet{Murray}, \citet{Mesinger11}, \citet{Park19}).\footnote{https://github.com/21cmfast/21cmFAST} Our data set consisted of 100 pairs of light-cones which differed only by whether the light-cones in the dataset were simulated using bubble-finding algorithms 1 or 2. We adopt a simulation size of 1 $\text{Gpc}^{3}$ and a $256^3$ voxel grid, with each simulation running from $z = 16.0$ to $z = 6.0$. Additionally, we vary two astrophysical parameters, $\zeta$ and $T_{vir}^{min}$, across our dataset in order to model reionization. $\zeta$ and $T_{vir}^{min}$ parameterize the photons emitted per ionizing source and the minimum mass of star-forming galaxies respectively. We also assume a spin temperature much larger than the CMB temperature, which eliminates any dependence on the X-ray emission parameters within 21cmFAST. We limited our main result to the parameterization described above in the interest of saving computing resources while simultaneously giving a strong qualitative description of 21CMMC's model dependence. In Section \ref{sec:algorithm_differences}, we outline the differences between bubble-finding algorithms 1 and 2, in Section \ref{sec:21cmFAST_setup}, we describe our parameterization of the EoR, and in Section \ref{sec:21CMMC_setup}, we describe our 21CMMC setup.

\subsection{Differences between Algorithms 1 and 2}
\label{sec:algorithm_differences}

As discussed in \citet{Mesinger11}, there is one difference between bubble-finding algorithms 1 and 2, which is manifest in their respective bubble filtering procedures. Beginning with density and velocity initial conditions in Lagrangian space, 21cmFAST generates an evolved density field by approximating gravitational collapse with first-order perturbation theory \citep{Zeldovich}. The ionization field is then generated from the evolved density field with the requirement that the number of ionizing photons in a region exceed the number of hydrogen atoms. Following \citet{HIIregions}, both algorithms then use the excursion-set approach \citep{Bond1991, Lacey93}, decreasing the filter scale from some maximum value $R_{\text{max}}$ down to the cell size, $R_{\text{cell}}$ in logarithmic steps with $R_{\text{next}} = 0.9$ $R_{\text{prev}}$. At each scale, both algorithms use their shared ionization criteria to determine whether the region in question is ionized. This ionization criteria is

\begin{equation}
    f_{coll} \geq {\zeta}^{-1}
\end{equation}
where $f_{coll}$ is the fraction of mass residing in collapsed halos within the bubble under consideration, and $\zeta$ is the ionizing efficiency of the galaxies.

The bubble-finding algorithms diverge in their treatment of a region which satisfies the ionization criteria. When a region meets the ionization threshold, bubble-finding algorithm 1 flags all pixels inside the region as ``ionized'', whereas bubble-finding algorithm 2 flags only the central pixel of the region. Bubble-finding algorithm 1 performs this filtering procedure on all pixels at all scales, regardless of whether the region in question overlaps a region which had previously been flagged as ionized. In contrast, by flagging pixel-by-pixel, bubble-finding algorithm 2's procedure does not permit overlap between ionized regions.

\begin{figure*}[ht!]
\includegraphics[trim = {3cm 0cm 3cm 1cm}, clip, width=\textwidth]{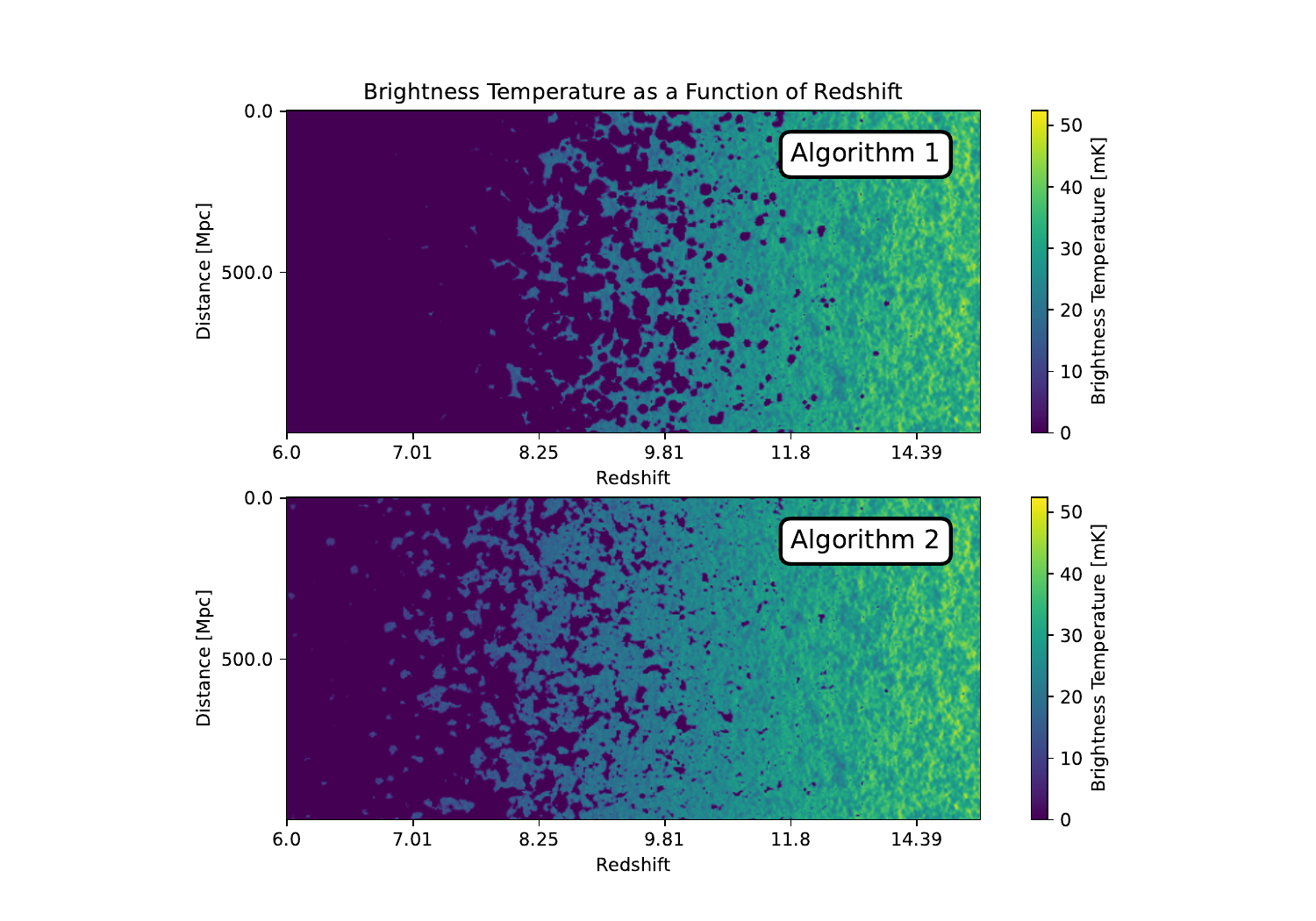}
\caption{Brightness temperature maps showing a slice along one spatial axis of a light-cone pair. The two light-cones differ only by whether they were simulated using bubble-finding algorithm 1 (top) or 2 (bottom). The underlying density field along with all astrophysical parameter values are equal between these two light-cones.}
\label{fig:lightcones}
\end{figure*}

This difference has a few major consequences for light-cones simulated using each algorithm. As discussed in Section \ref{sec:introduction}, reionization is thought to have progressed through ionized bubbles around luminous sources expanding outwards and overlapping. Thus, early work suggested that bubble-finding algorithm 1 was the more physically accurate of the two; \citet{Mesinger11} showed that algorithm 1 better reproduced the HII morphological structure of an RT simulation than did algorithm 2. This point has come under contention in recent years, with \citet{Hutter} finding that bubble-finding algorithm 2 resulted in better agreement with a full RT simulation. However, the question of which algorithm is the more physically accurate is not pertinent to this analysis, as both algorithms  represent plausible morphologies for the EoR and our analysis focuses on testing whether 21CMMC can give reasonable results when the underlying model cannot exactly reproduce the data (as will be the case for real data from the real universe).

Simulations using bubble-finding algorithm 1 are more computationally expensive than simulations using bubble-finding algorithm 2. Bubble-finding algorithm 1 approaches $O(N^{2})$ at low redshifts where $N$ is the number of grid cells, while bubble-finding algorithm 2 is $O(N)$. Due to being less computationally expensive, bubble-finding algorithm 2 is the default algorithm in 21cmFAST. On a single core, generating one 1 $\text{Gpc}^3$ light-cone with a $256^3$ voxel grid from z = 16 to 6 takes $\approx 45$ minutes with algorithm 2 and $\approx 4$ hours with algorithm 1.

\begin{figure*}[ht!]
\includegraphics[trim = {0cm 0cm 0cm 0cm}, clip, width=\textwidth]{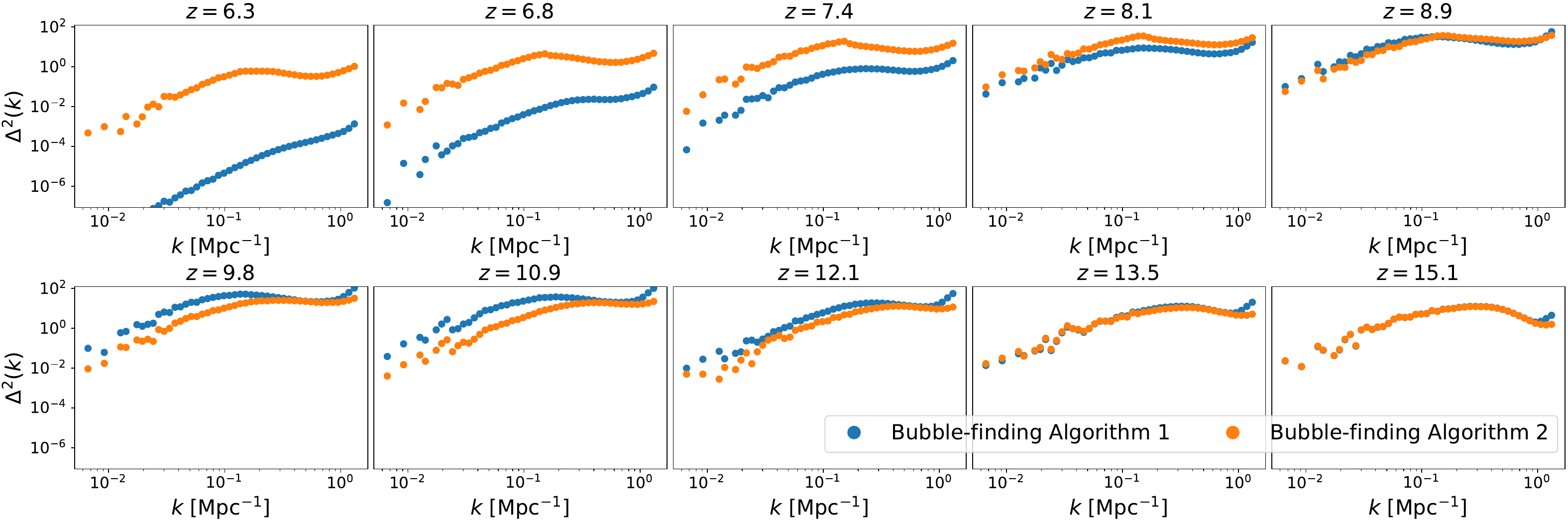}
\caption{Comparison of the 21 cm PS of a bubble-finding algorithm 1 and 2 light-cone pair. This light-cone pair corresponds to that used in Figure \ref{fig:lightcones}. Both light-cones were simulated with the same density field.}
\label{fig:ps-compare}
\end{figure*}

The difference in filtering procedures also has consequences on the topology of reionization. Aesthetically, \citet{Mesinger07} showed that brightness temperature maps simulated using bubble-finding algorithm 1 are more ``bubbly'' than maps simulated using bubble-finding algorithm 2, meaning that light-cones simulated with the same input parameters will look visibly different if run on different algorithms. This is demonstrated in Figure \ref{fig:lightcones}, which shows two brightness-temperature maps simulated using the same input parameters and underlying density field, differing only by which bubble-finding algorithm was used in the simulation. Ionized regions in the top panel (bubble-finding algorithm 1) have visibly less small-scale structure than in the bottom panel (bubble-finding algorithm 2) and look as if they were constructed by overlaying ionized circles atop each other. Of more consequence for this analysis, by flagging the entire sphere when performing the filtering procedure, bubble-finding algorithm 1 ionizes \textit{at least} as many pixels per time step as bubble-finding algorithm 2. In other words, reionization in bubble-finding algorithm 1 simulations progresses \textit{at least} as fast as in bubble-finding algorithm 2. While visually apparent in Figure \ref{fig:lightcones}, the difference in reionization speed also has a significant effect on the PS. Figure \ref{fig:ps-compare} shows the PS for the same bubble-finding algorithm 1 and 2 light-cone pair (blue vs orange points respectively) shown in Figure \ref{fig:lightcones}. By $z = 8.1$, the bubble-finding algorithm 1 PS has less power than the bubble-finding algorithm 2 PS, and by $z=6.3$, the bubble-finding algorithm 1 signal is more than three orders of magnitude weaker than that of bubble-finding algorithm 2.  This dramatic decrease in power is due to the disappearance of neutral hydrogen from the IGM at the tail-end of reionization.

\subsection{21cmFAST Setup}
\label{sec:21cmFAST_setup}

We vary two astrophysical parameters in this paper in order to construct a sufficiently robust data set encapsulating the diversity of plausible models of the EoR. Below we describe the parameters we included in our model. We emphasize that the parameter ranges described in Sections \ref{sec:ionizing_efficiency} - \ref{sec:virial_temp} correspond to those used to create our set of input light-cones; the prior ranges used in 21CMMC towards fitting these light-cones are described in Section \ref{sec:21CMMC_setup}.

\subsubsection{Ionizing Efficiency ($\zeta$)}
\label{sec:ionizing_efficiency}

$\zeta$, or the ionizing efficiency of high-z galaxies is defined as

\begin{equation}
    \zeta = 30\left(\frac{f_{esc}}{0.2}\right)\left(\frac{f_{*}}{0.05}\right)\left(\frac{N_{\gamma}}{4400}\right)\left(\frac{1.5}{1+n_{rec}}\right)
\end{equation}
with $f_{esc}$ the fraction of ionizing photons escaping into the IGM, $f_{*}$ the fraction of galactic gas in stars, $N_{\gamma}$ the number of ionizing photons produced per baryon in stars, and $n_{rec}$ the average number of recombinations per baryon in the IGM. $\zeta$ is inversely correlated to the duration of reionization. In this work, we assigned randomly selected values to our light-cones in the range $\zeta \in [5,100]$. This range is consistent with that used in \citet{GreigMes15}.

\subsubsection{Minimum Virial Temperature ($T_{vir}^{min}$)}
\label{sec:virial_temp}

The minimum mass threshold for a halo to host a star-forming galaxy can be defined in terms of its virial temperature, $T_{vir}^{min}$, which regulates processes important to star formation such as gas accretion, cooling, and retainment of supernovae outflows. $T_{vir}^{min}$ can be related to its halo mass via 

\begin{equation}
\begin{split}
M_{\min} = 10^{8} h^{-1}
    \left(\frac{\mu}{0.6}\right)^{-3/2}
    \left(\frac{\Omega_m}{\Omega_m^{z}}
          \frac{\Delta_c}{18\pi^{2}}\right)^{-1/2} \\
    \times
    \left(\frac{T_{\rm vir}}{1.98\times10^{4}\,\mathrm{K}}\right)^{3/2}
    \left(\frac{1+z}{10}\right)^{-3/2}
    M_{\odot}.
\end{split}
\end{equation}

\noindent\citep{GreigMes15} with $\mu$ the mean molecular weight, $\Omega_{m}$ the matter density parameter, $\Omega_{m}^{z} = \Omega_{m}(1+z)^{3}/[\Omega_{m}(1+z)^{3}+ \Omega_{\Lambda}]$ with $\Omega_{\Lambda}$ the dark energy density parameter, and $\Delta_{c} = 18\pi^{2}+82d-39d^{2}$ where $d = \Omega_{m}^{z}-1$. $T_{vir}^{min} \approx10^4\text{ K}$ corresponds to the minimum threshold temperature for efficient atomic cooling, although efficient star formation likely occurs at higher values \citep{Springel03}.

$T_{vir}^{min}$ influences the timing of reionization and the bias of the galaxies responsible for reionization. For a fixed mean neutral fraction, a higher value of $T_{vir}^{min}$ corresponds to a later (lower $z$) reionization with more large-scale structure. Due to its effect on the bias of galaxies responsible for reionization, $T_{vir}^{min}$ is also inversely correlated to the duration of reionization, with higher values of $T_{vir}^{min}$ corresponding to shorter reionization models with brighter ionizing sources. In this work, we assigned randomly selected values to our light-cones in the range $\log(T_{vir}^{min}) \in [4.0,6.0]$. This range is consistent with the range published in \citet{GreigMes17}.

\subsection{21CMMC Setup}
\label{sec:21CMMC_setup}

21CMMC is a publicly available MCMC driver of 21cmFAST \citep{GreigMes15,GreigMes17,GreigMes18,Park19}.\footnote{https://github.com/21cmfast/21CMMC} In this work, we use 21CMMC to fit the 21 cm PS of the light-cones generated as described in Section \ref{sec:21cmFAST_setup}. The 21 cm PS was extracted from the light-cones in our dataset by slicing each light-cone into ten ``chunks'', with each chunk spanning an equivalent comoving volume, and then calculating the 2D PS for each chunk individually. The MCMC procedure generates 3D realizations of the 21 cm signal with a box length of 250 $\text{Mpc}^{3}$ and $64^3$ voxels, and we restricted 21CMMC to fit only $k$-modes $0.1\leq k \leq 1.0$. Additionally, we assume a 15$\%$ modeling uncertainty on our input PS, which is the default setting of 21CMMC. We discuss modeling uncertainty and its impact on this work in Section \ref{sec:degeneracy-explanation}. For a technical discussion of 21CMMC, we refer the reader to \citet{GreigMes15}.

In this work, we sampled only $T_{vir}^{min}$ and $\zeta$. We limited 21CMMC to sampling only two parameters for two reasons. First, sampling additional parameters would require additional computing resources for our runs of 21CMMC. Secondly, limiting the number of parameters for 21CMMC to fit should, in principle, improve the accuracy of its returns. If 21CMMC fails to fit two parameters, we should be skeptical of its fits using more complicated parameterizations of the EoR. We performed a cursory analysis using the more modern and flexible \citet{Park19} EoR parameterization, which encompasses the parameters $f_{*,10}$, $\alpha_{*}$, $f_{esc,10}$, $\alpha_{esc}$, $t_{*}$, and $M_{turn}$. In this analysis, we provided 21CMMC, at the outset of each run, with the fiducial values of $f_{*,10}$ and $\alpha_{*}$ in order to avoid their degeneracy with $f_{esc,10}$ and $\alpha_{esc}$. We thus sampled $f_{esc,10}$, $\alpha_{esc}$, $t_{*}$, and $M_{turn}$, and found that our results were qualitatively unchanged from those of our main parameterization, with the added drawback of being more computationally expensive. For a detailed description of each parameter mentioned above, see \citet{Park19}.

We assume flat priors on both $\log(T_{vir}^{min})$ and $\zeta$ and allow 21CMMC to sample in the ranges $\log(T_{vir}^{min}) \in [3.5,6.5]$ and $\zeta \in [5,100]$. We allowed 21CMMC to sample $T_{vir}^{min}$ beyond its allowed parameter range because doing so did not meaningfully affect the accuracy of its returns; however, as we shall demonstrate, even confining the priors to the allowed values of $\zeta$ was insufficient to yield accurate results. Each run was given 200 sampling iterations, consistent with \citet{GreigMes17}.  200 iterations was sufficient to observe convergence for most light-cones, with tests of up to 1000 total iterations failing to make any significant changes to the recovered parameters.  Because of the computational expense, we use only six walkers to explore the parameter space, which is in agreement with the default value of ``walkersRatio'' in 21CMMC. The results in \citet{GreigMes17} come from using 400 walkers, although other works have shown that significantly fewer than 400 walkers are needed to accurately constrain the astrophysical parameterization of the EoR \citep{Gazagnes, Nasirudin, Binnie, GreigMes18}. The results of a preliminary analysis we conducted with 400 walkers did not meaningfully differ from the results described in this work. All runs were carried out over the full range of redshifts for which the input 21 cm PS was nonzero.

\section{Results} \label{sec:floats}
\label{sec:results}

In this section, we present our main result, 21CMMC's returned $\zeta$ and $T_{vir}^{min}$ values for \emph{in-domain} (i.e. algorithmic agreement between 21CMMC and the input data) and \emph{out-of-domain} runs (i.e. algorithmic disagreement between 21CMMC and the input data).
While these parameters are physically-motivated, they are not exact quantities with ``true'' values to be measured in our Universe.  Moreover, they are specific to the 21cmFAST algorithm and do not necessarily have well-defined values even in other simulations.  We choose to focus on them primarily because analyses of real data have also focused on them: recent papers interpreting LOFAR \citep{GreigB}, MWA \citep{GreigA}, and HERA \citep{HeraA} all contain posterior distributions for the parameter values recovered from 21CMMC, while \citet{HeraB} focuses extensively on their posterior constraints on 21cmFAST's X-ray parameters.  Given the value the community places on the physical interpretation of these parameters, it becomes important to know whether these constraints have an underlying model dependence.  In Section \ref{sec:observables} we present an additional analysis focused on the recovery of parameters describing the ionization history to see if these nominally model-independent parameters can be recovered better than 21cmFAST-specific parameters studied here.

In examining 21CMMC's performance on our two-parameter model, it is illustrative to examine the sampler's performance in recovering each parameter separately. Our results fall into one of four conditions: an algorithm 1 light-cone with an algorithm 1 sampler, an algorithm 2 light-cone with an algorithm 2 sampler, an algorithm 1 light-cone with an algorithm 2 sampler, and an algorithm 2 light-cone with an algorithm 1 sampler. Going forward, these conditions will be referred to as conditions 1$\rightarrow$1, 2$\rightarrow$2, 1$\rightarrow$2, and 2$\rightarrow$1 respectively (i.e. the first number refers to the bubble-finding algorithm used to generate the input light-cone, and the second number refers to the bubble-finding algorithm used in the 21CMMC fits to the input light-cone).

\subsection{$\zeta$ Recovery}

\begin{deluxetable*}{rlllll}
\digitalasset
\tablecaption{Main Result $\zeta$ and $T_{vir}^{min}$ $\chi_{\nu}^{2}$ Values \label{tab:description}}
\tablehead{
\colhead{Condition Name} & \colhead{21cmFAST Algorithm} & \colhead{21CMMC Algorithm} & \colhead{$\zeta$ $\chi^{2}_{\nu}$} & \colhead{$T_{vir}^{min}$ $\chi^{2}_{\nu}$}
}
\startdata
$1\rightarrow1$ & Bubble-Finding Algorithm 1 & Bubble-Finding Algorithm 1 & 22.092 & 7.649 \\
$2\rightarrow2$ & Bubble-Finding Algorithm 2 & Bubble-Finding Algorithm 2 & 6.417 & 2.608  \\
$1\rightarrow2$ & Bubble-Finding Algorithm 1 & Bubble-Finding Algorithm 2 & 42.427 & 78.180  \\
$2\rightarrow1$ & Bubble-Finding Algorithm 2 & Bubble-Finding Algorithm 1 & 40.083 & 10.964 \\
\enddata
\tablecomments{Main result $\chi^{2}_{\nu}$ values for $\zeta$ and $T_{vir}^{min}$ returns for conditions 1$\rightarrow$1 (top), 2$\rightarrow$2 (second from the top), 1$\rightarrow$2 (second from the bottom), and 2$\rightarrow$1 (bottom).}
\label{tab:tab_1}
\end{deluxetable*}

\begin{figure*}[ht!]
\includegraphics[trim = {0cm 0cm 0cm 0cm}, clip, width=\textwidth]{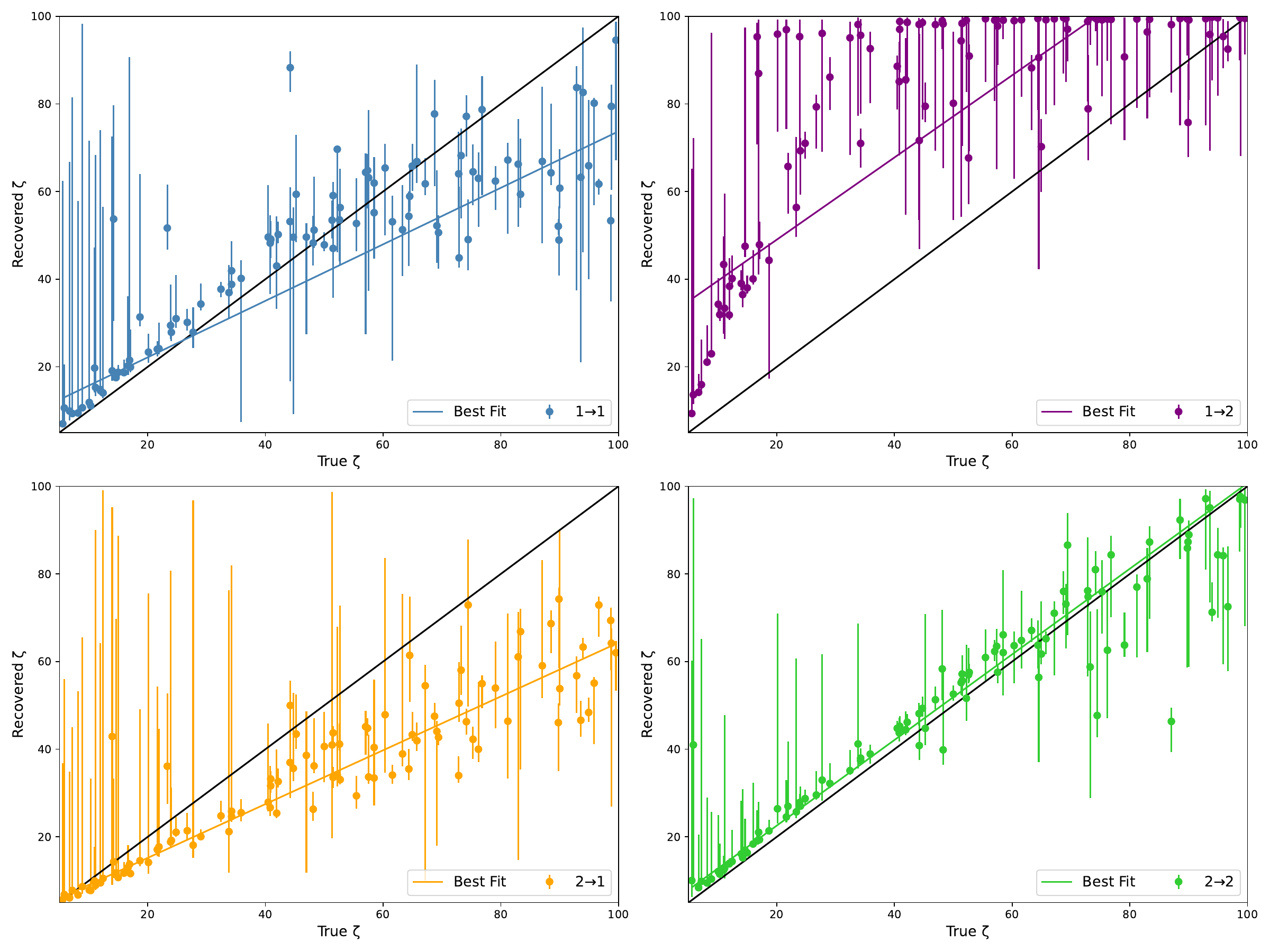}
\caption{Returned $\zeta$ values for conditions 1$\rightarrow$1 (top left), 2$\rightarrow$2 (bottom right), 1$\rightarrow$2 (top right), and 2$\rightarrow$1 (bottom left). The black line in each subplot represents a perfect $\zeta$ recovery (i.e. y=x), and the solid colored line in each subplot represents the line of best fit for our recovered $\zeta$ values.}
\label{fig:main-zeta}
\end{figure*}

In Figure \ref{fig:main-zeta}, we present 21CMMC's recovery of $\zeta$ for conditions 1$\rightarrow$1, 2$\rightarrow$2, 1$\rightarrow$2, and 2$\rightarrow$1. We plot the recovered value of $\zeta$ for each run against the true-value for the light-cone. The error bars represent 21CMMC's 68$\%$ confidence interval for each returned $\zeta$ value, and the black line in each subplot represents a perfect $\zeta$ recovery. The solid colored line in each subplot represents the line of best fit for our recovered $\zeta$ values. These results are also summarized in Table \ref{tab:tab_1}, which presents the $\chi^2_\nu$ for the returned results compared with the true values for both $\zeta$ and $T_{vir}^{min}.$

One important observation we can make from Figure \ref{fig:main-zeta} is that condition 2$\rightarrow$2 (bottom right) tightly follows the identity line for the entire allowed range of $\zeta$ (albeit with a slight positive bias). This result --- in-domain performance with bubble algorithm 2 --- has been established by existing literature on 21CMMC \citep{GreigMes15}. Condition 1$\rightarrow$1 (top left), meanwhile, broadly mirrors condition 2$\rightarrow$2 with the exception that condition 1$\rightarrow$1 returns higher values of $\zeta$ less accurately than condition 2$\rightarrow$2. In fact, we can see that for $\zeta \gtrsim 60$, 21CMMC's outputs appear to level off and diverge from the identity line. Uncertainties on the returned $\zeta$ values for condition 1$\rightarrow$1 are also wider in general than for condition 2$\rightarrow$2.

Condition 1$\rightarrow$2 (top right) reveals that 21CMMC run on bubble-finding algorithm 1 light-cones categorically overestimates $\zeta$ when sampling bubble-finding algorithm 2. Also worth noting for condition 1$\rightarrow$2 is that it appears to follow a linear trend beneath true $\zeta \lesssim 30$, at which point 21CMMC's returns pile up at the ceiling of the allowed range. This suggests that there exists some constant of proportionality between the speed of reionization in bubble-finding algorithm 1 light-cones and bubble-finding algorithm 2 light-cones. We leave the determination of this relationship to future work. The quoted $\chi_{\nu}^{2}$ value for condition 1$\rightarrow$2, shown in Table \ref{tab:tab_1}, is aided by the ceiling effect: had 21CMMC been given a broader range of $\zeta$ to explore, this value would no doubt well exceed its current value. This assertion is supported by a cursory analysis we performed in which we allowed 21CMMC to sample $\zeta$ in the range $\zeta \in [0,200]$, and indeed found that the $\chi_{\nu}^{2}$ value for condition 1$\rightarrow$2 increased. 

Condition 2$\rightarrow$1 (bottom left) shows a complementary result to condition 1$\rightarrow$2. 21CMMC run on bubble-finding algorithm 2 light-cones categorically underestimates $\zeta$ when sampling bubble-finding algorithm 1. Similarly to condition 1$\rightarrow$2, too, condition 2$\rightarrow$1 appears to follow a linear trend, which is not obscured by the ceiling effect. This, too, suggests a constant of proportionality between the speed of reionization in bubble-finding algorithm 1 and 2 light-cones. As mentioned above, we leave exploration of this result to future work.

The most important thing to notice from Figure \ref{fig:main-zeta} is that condition 1$\rightarrow$1 outperforms condition 2$\rightarrow$1, and condition 2$\rightarrow$2 outperforms condition 1$\rightarrow$2. Figure \ref{fig:main-zeta} demonstrates that 21CMMC has a clear model dependence between the mock data and the sampler in returning $\zeta$.  The significantly higher $\chi^2_\nu$ values shown in Table \ref{tab:tab_1} for the out-of-domain versus in-domain analyses also support this conclusion.

\subsection{$T_{vir}^{min}$ Recovery}

\begin{figure*}[ht!]
\includegraphics[trim = {0cm 0cm 0cm 0cm}, clip, width=\textwidth]{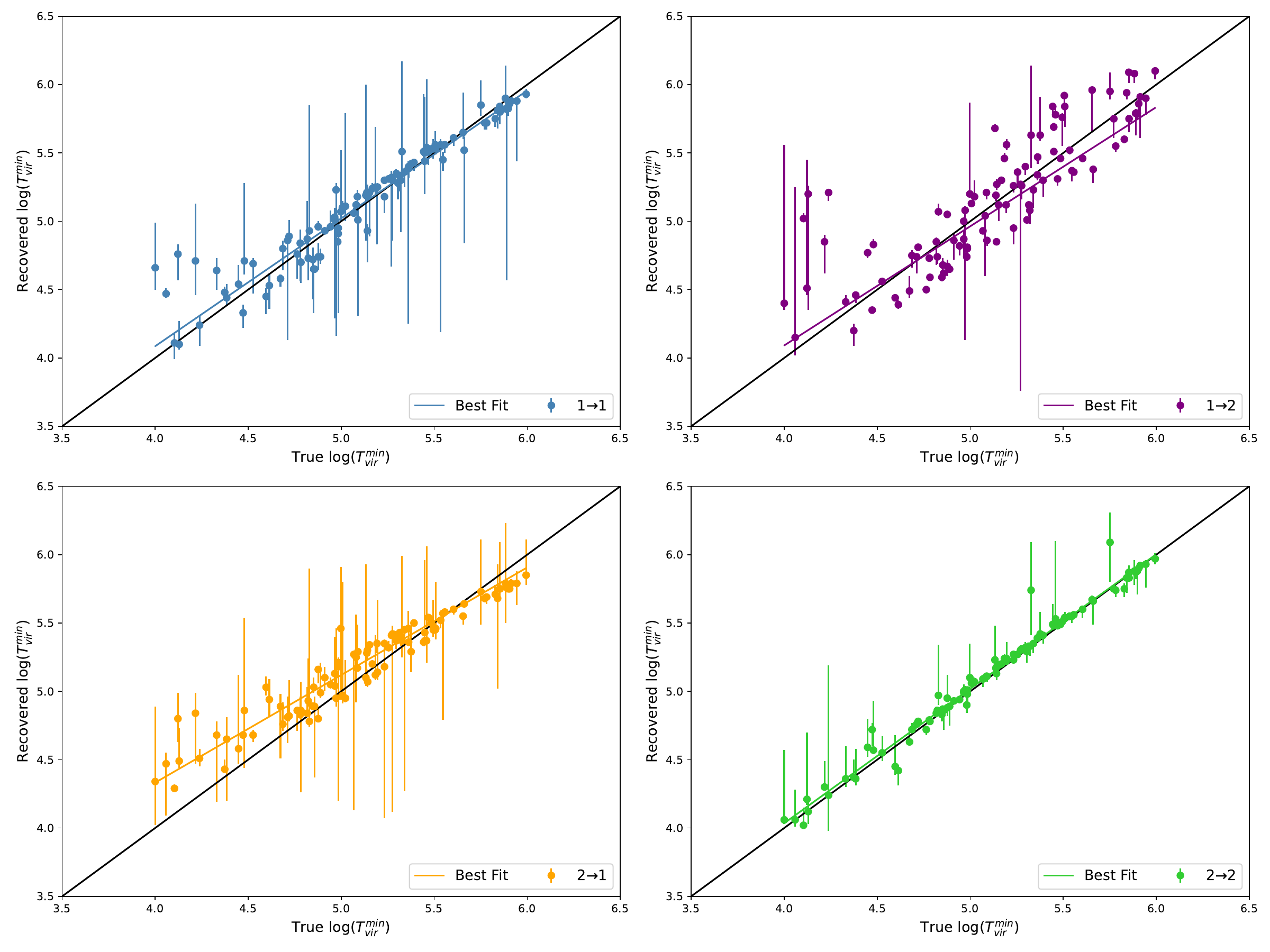}
\caption{Returned $T_{vir}^{min}$ values for conditions 1$\rightarrow$1 (top left), 2$\rightarrow$2 (bottom right), 1$\rightarrow$2 (top right), and 2$\rightarrow$1 (bottom left). The black line in each subplot represents a perfect $T_{vir}^{min}$ recovery (i.e. y=x), and the solid colored line in each subplot represents the line of best fit for our recovered $T_{vir}^{min}$ values. One anomalous return in condition 1$\rightarrow$2 with no asymmetric error was omitted from the calculation of our line of best fit.}
\label{fig:main-tvir}
\end{figure*}

In Figure \ref{fig:main-tvir}, we present 21CMMC's performance in recovering $T_{vir}^{min}$ values for conditions 1$\rightarrow$1, 2$\rightarrow$2, 1$\rightarrow$2, and 2$\rightarrow$1. Again, we plot true $T_{vir}^{min}$ values against recovered $T_{vir}^{min}$ values for each light-cone, with error bars representing 21CMMC's 68$\%$ confidence intervals, the black line in each subplot representing a perfect $T_{vir}^{min}$ return (i.e. y=x), and the solid colored line in each subplot representing the line of best fit for our returned $T_{vir}^{min}$ values.

Conditions 1$\rightarrow$1 and 2$\rightarrow$2 (top left and bottom right) are broadly similar to the results presented in the previous section. 21CMMC performs well when there exists algorithmic agreement between the mock data and the sampler.

While the line of best fit for condition 1$\rightarrow$2 (top right) closely follows the identity line, the returned values are distributed both above and below the identity line. The $\chi_{\nu}^{2}$ value in this case is a much better indication of 21CMMC's performance, which reflects that condition 2$\rightarrow$2's results $(\chi_{\nu}^{2} = 2.608)$ are much more tightly clustered about the identity line than condition 1$\rightarrow$2 $(\chi_{\nu}^{2} = 78.180)$. Again, the $\chi_{\nu}^{2}$ values for each condition are displayed in Table \ref{tab:tab_1}. 21CMMC outputted one anomalous $T_{vir}^{min}$ return for condition $1\rightarrow2$ with no error bar. This return was omitted in calculating the $\chi_{\nu}^{2}$ value as well as the line of best fit.

In contrast, when run on a bubble-finding algorithm 2 light-cone, 21CMMC, sampling bubble-finding algorithm 1, is still able to return values of $T_{vir}^{min}$ which tightly fit the identity line. Note, however, that condition 1$\rightarrow$1 $(\chi_{\nu}^{2} = 7.649)$ does outperform condition 2$\rightarrow$1 $(\chi_{\nu}^{2} = 10.964)$. This suggests that in returning values of $T_{vir}^{min}$, 21CMMC still suffers a model dependence between the mock data and the sampler, albeit a much weaker dependence than in the case of $\zeta$. We propose a possible explanation in Section \ref{sec:degeneracy-explanation}.

\section{Recreating a Bubble-Finding Algorithm 1 PS}
\label{sec:PS_recreate}

In this section, we present a follow-up test we conducted in order to determine why 21CMMC exhibits a stronger model dependence in sampling $\zeta$ than $T_{vir}^{min}$. We conducted three runs of 21CMMC with condition 1$\rightarrow$2 to this end. In our first run, we allowed 21CMMC to sample both $\zeta$ and $T_{vir}^{min}$, and in addition, we expanded our allowed $\zeta$ range from $\zeta \in [5,100]$ to $\zeta \in [5,200]$ (hereafter referred to as Run B for ``both''). We adjusted the initial $\zeta$ guess of our run to the midpoint of our new range, and again allowed 21CMMC to sample for 200 iterations. We expanded our allowed $\zeta$ range so as to avoid the ceiling effect common in other 21CMMC runs of condition 1$\rightarrow$2. In our second run, we provided 21CMMC with the true $\zeta$ value at the outset of the run and sampled only $T_{vir}^{min}$ (Run T), and in our third run, we provided 21CMMC with the true $T_{vir}^{min}$ value at the outset of the run and sampled only $\zeta$, again with an expanded range (Run Z). 21CMMC's returned parameter values are displayed in Table \ref{tab:tab_2}.

\begin{figure*}[ht!]
\includegraphics[trim = {0cm 0cm 0cm 0cm}, clip, width=\textwidth]{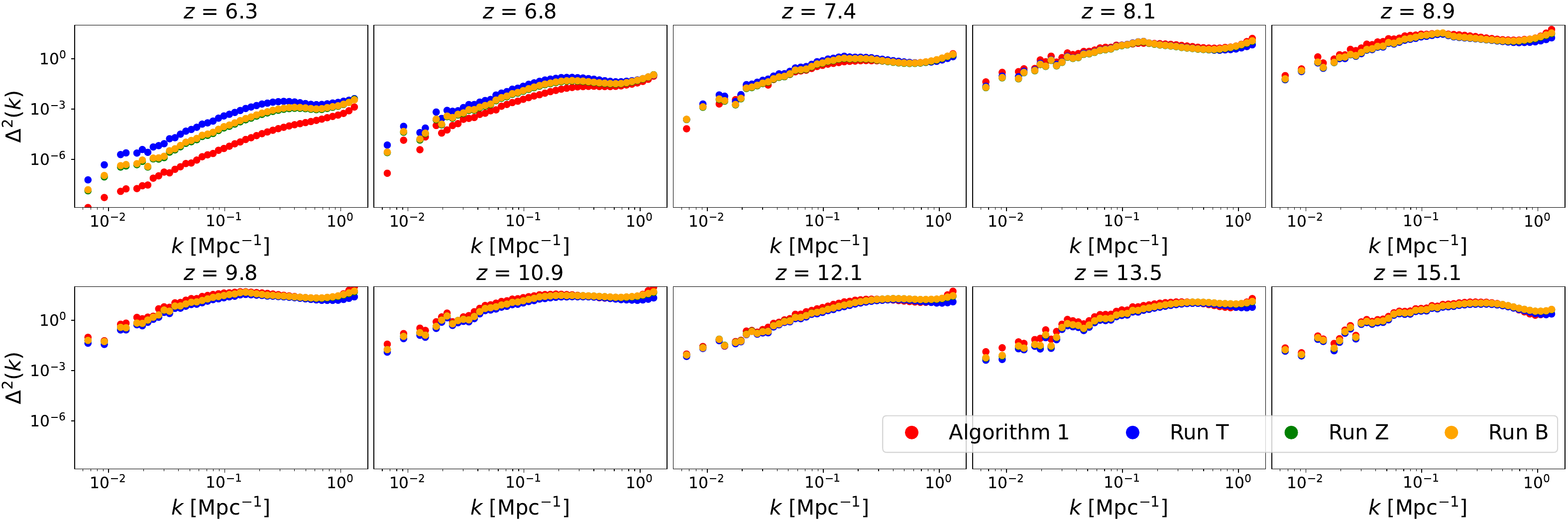}
\caption{Comparison of the 21 cm PS for a light-cone simulated using bubble-finding algorithm 1 (red), bubble-finding algorithm 2 with the median parameters of a run of 21CMMC with an expanded $\zeta$ range (Run B, yellow), bubble-finding algorithm 2 varying only $T_{vir}^{min}$ (Run T, blue), and bubble-finding algorithm 2 varying only $\zeta$ with an expanded range (Run Z, green).}
\label{fig:PS_recreate}
\end{figure*}

After completing our runs of 21CMMC, we used 21cmFAST to simulate bubble-finding algorithm 2 light-cones with the median values of 21CMMC's outputted parameters, with all other astrophysical parameters equal to the bubble-finding algorithm 1 light-cone. In Figure \ref{fig:PS_recreate}, we present a comparison of the 21 cm PS for the four light-cones described above. 

It is important to note that while the PS for all four light-cones are qualitatively similar, the light-cone corresponding to Run T (blue) is the most dissimilar from the bubble-finding algorithm 1 light-cone (red). Run T tends to disagree from the other three light-cones somewhat at small scales (large $k$), and ionizes slower than the other three. This result suggests that the difference between bubble-finding algorithms 1 and 2 is degenerate with $\zeta$, and not that some property of $T_{vir}^{min}$ allows 21CMMC to return it accurately in out-of-domain runs. As shown in Table \ref{tab:tab_2}, 21CMMC returned similar values of $\zeta$ in both Runs B and Z, supported further by the fact that the PS corresponding to Runs B and Z (yellow and green) are the most similar of the four PS shown in Figure \ref{fig:PS_recreate}. Furthermore, 21CMMC accurately returned $T_{vir}^{min}$ in Run B, but when sampling $T_{vir}^{min}$ alone, 21CMMC returned an inaccurate value of $T_{vir}^{min}$ in order to overlap the ionization history of bubble-finding algorithm 1 light-cones. This suggests that the difference between the two bubble-finding algorithms is better characterized by a change in $\zeta$ than by a change in $T_{vir}^{min}$, but in the absence of the ability to sample $\zeta$, 21CMMC adjusts $T_{vir}^{min}$ in a compromise between affecting the shape of the PS at high redshifts and having an overlapping ionization history at low redshifts, leading to a PS which is the most dissimilar from the bubble-finding algorithm 1 light-cone of the light-cones shown in Figure \ref{fig:PS_recreate}.

\setlength{\tabcolsep}{24pt}
\begin{deluxetable*}{rll}
\digitalasset
\tablecaption{Returned $\zeta$ and $T_{vir}^{min}$ Values \label{tab:description}}
\tablehead{
\colhead{} & \colhead{$\zeta$} & \colhead{$\log(T_{vir}^{min})$}
}
\startdata
Fiducial & 63.30 & 4.82 \\
Run B & $137.98^{+6.01}_{-20.64}$ & $4.85^{+0.02}_{-0.47}$ \\
Run T & N/A & $4.54^{+0.01}_{-0.00}$ \\
Run Z & $129.56^{+0.39}_{-0.37}$ & N/A \\
\enddata
\tablecomments{Returned $\zeta$ and $T_{vir}^{min}$ values for runs of 21CMMC with an expanded $\zeta$ range (second from the top; Run B), varying only $T_{vir}^{min}$ (second from the bottom; Run T), and varying only $\zeta$ with an expanded range (bottom; Run Z).}
\end{deluxetable*}
\label{tab:tab_2}

This test shows that the relative accuracy with which 21CMMC returns values of $T_{vir}^{min}$ is a product of the nature of the difference between the two bubble-finding algorithms and our choice of the parameters we varied in this work. $T_{vir}^{min}$ is not a parameter which is guaranteed to be accurately returned in any out-of-domain run of 21CMMC. Instead, the specific difference between the two bubble-finding algorithms in this work is better characterized by a change in $\zeta$ than by a change in $T_{vir}^{min}$. Had we chosen different simulation algorithms of the EoR or chosen to sample $T_{vir}^{min}$ alongside some other parameter than $\zeta$, for instance a parameter uncorrelated with the duration of reionization, 21CMMC would not necessarily have accurately returned $T_{vir}^{min}$. We discuss why the difference between bubble-finding algorithms 1 and 2 is degenerate with $\zeta$ in Section \ref{sec:degeneracy-explanation}

\section{Returning Model-Independent Parameters}
\label{sec:observables}

In this section, we discuss the ramifications of this work on experimental EoR research as well as the impact of 21CMMC's model dependence on the measurement of model-independent observables of the EoR.

They key question is whether the results of this analysis are limited to parameters like $\zeta$ and $T_{vir}^{min}$. We expect 21CMMC to exhibit a model dependence in returning these parameters because these parameters are specific to 21cmFAST. This issue could, in principle, be avoided if researchers did not use 21CMMC to assign 21cmFAST parameters to the Universe. Instead, 21CMMC can be used to constrain more phenomenological parameters. Whereas the Universe does not have fiducial values of $\zeta$ and $T_{vir}^{min}$, it does have an ionization history, and researchers can utilize 21CMMC towards constraining, among other parameters, the duration and midpoint of reionization (and, indeed, the aforementioned analyses of \citealt{GreigB}, \citealt{GreigA}, and \citealt{HeraA} also derive constraints on ionization histories through 21CMMC.)

Here we investigate whether 21CMMC can accurately return the midpoint and duration of reionization in out-of-domain runs despite the failure to recover $\zeta$ documented in Section \ref{sec:results}. We use the returned maximum likelihood $\zeta$ and $T_{vir}^{min}$ values from our runs of 21CMMC under condition $2\rightarrow1$ to simulate new bubble-finding algorithm 1 light-cones. We then compared the midpoint and duration of reionization of our new light-cones to the original bubble-finding algorithm 2 light-cones. We defined the duration of reionization as the difference between the redshifts at which the Universe was 75$\%$ neutral and 25$\%$ neutral. While it is in principle possible to propagate uncertainties on $\zeta$ and $T_{vir}^{min}$ to the midpoint and duration of reionization, we focus here only on the ionization histories yielded by the maximum a posteriori values of these parameters, which we find sufficient to illustrate the effect.

In our analysis of the duration of reionization, we excluded any light-cone pair for whom the IGM was more than 25$\%$ neutral at $z = 6.0$ for at least one light-cone in the pair. In our analysis of the midpoint of reionization, we excluded any light-cone pair for whom the IGM was more than 50$\%$ neutral at $z = 6.0$ for at least one light-cone in the pair.

We limited this analysis to our condition $2\rightarrow1$ runs of 21CMMC because our condition $1\rightarrow2$ runs suffered from the ceiling effect. We did, however, conduct a similar analysis as described above under condition $1\rightarrow2$ after performing runs of 21CMMC with an expanded $\zeta$ range, which yielded very similar results.

\begin{figure*}[ht!]
\includegraphics[trim = {0cm 0cm 0cm 0cm}, clip, width=\textwidth]{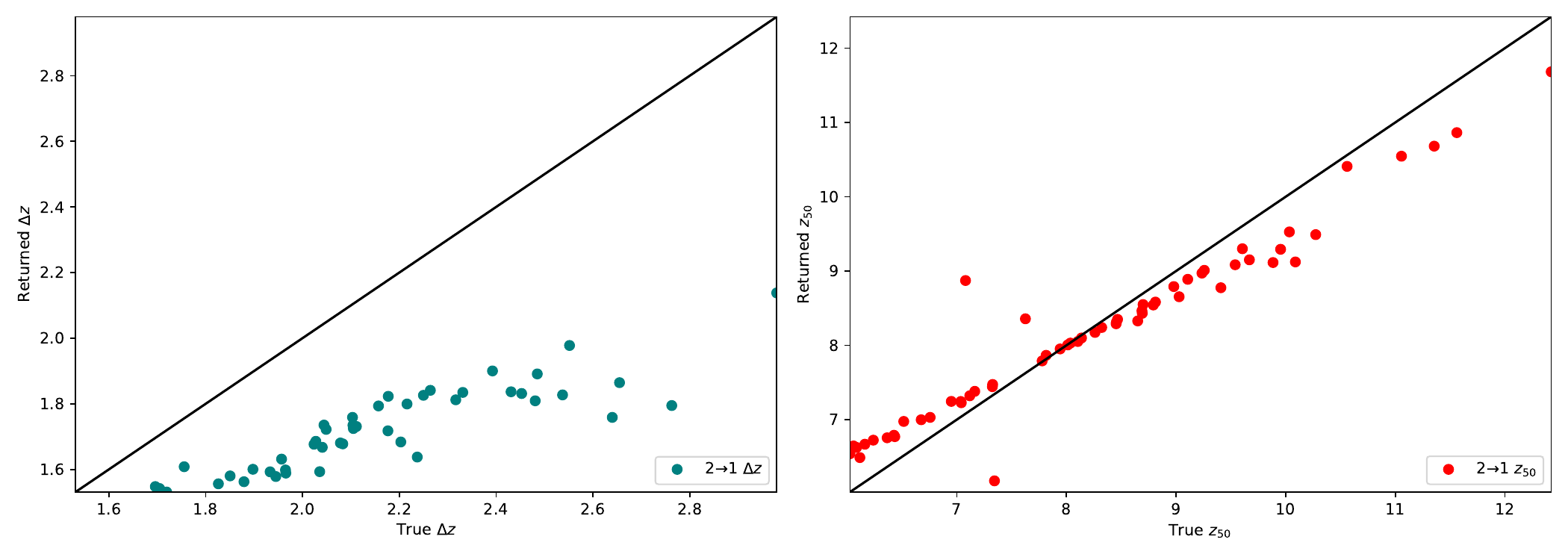}
\caption{21CMMC's returned durations of reionization (left) and midpoints of reionization (right) for condition $2\rightarrow1$ light-cones. The black line in each subplot represents a perfect recovery of duration or midpoint (i.e. y=x). Light-cones pairs for whom at least one light-cone was more than $25\%$ neutral at $z = 6.0$ or more than $50\%$ neutral at $z = 6.0$ were excluded from our plots of duration and midpoint of reionization, respectively.}
\label{fig:observables}
\end{figure*}

Figure \ref{fig:observables} shows the duration and midpoint of reionzation for the bubble-finding algorithm 1 light-cones simulated as described above plotted against the duration and midpoint of reionization for their corresponding bubble-finding algorithm 2 light-cones.

The left panel of Figure \ref{fig:observables} shows that 21CMMC is unable to accurately return the duration of reionization in out-of-domain runs. The bubble-finding algorithm 1 light-cones ionized much quicker than did their bubble-finding algorithm 2 counterparts. Although our analysis did not propagate uncertainty into our returned durations, the returned durations were clearly biased towards low values.

The right panel of Figure \ref{fig:observables}, meanwhile, shows that 21CMMC largely recovers the midpoint of reionization. The returned midpoints are tightly clustered around the identity line, and it is possible that were uncertainty properly accounted for, 21CMMC's returned midpoints would be consistent with the input bubble-finding algorithm 2 light-cones.

This result demonstrates that 21CMMC's model dependence prevents it from returning model-independent parameters as well. The implications of this result will be discussed further in Section \ref{sec:discussion}.

\section{\text{Incorporating Instrumental Noise}}
\label{sec:instrumental_noise}

\begin{figure*}[ht!]
\includegraphics[trim = {0cm 0cm 0cm 0cm}, clip, width=\textwidth]{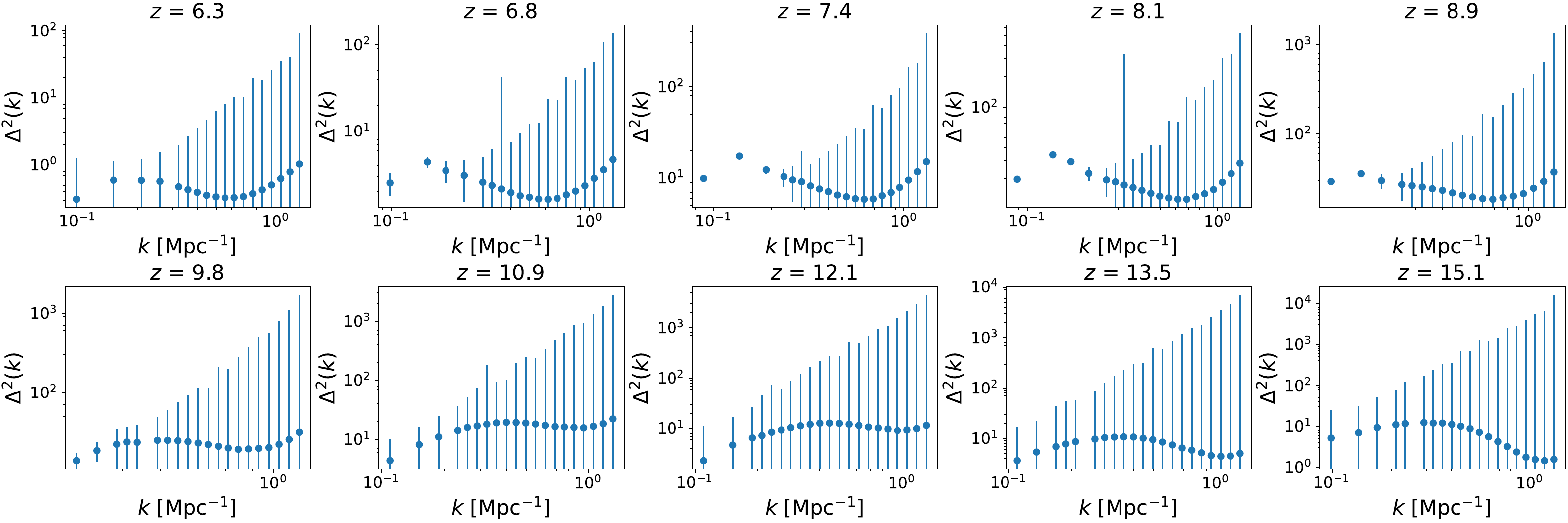}
\caption{Assumed uncertainties for our 21cmSense ``moderate'' noise model. $k$-modes for which 21cmSense outputted infinite uncertainty were excised from our power spectra files. This light-cone corresponds to the bubble-finding algorithm 2 light-cone used in Figure \ref{fig:lightcones}.}
\label{fig:noise_ps}
\end{figure*}

\begin{figure*}[ht!]
\includegraphics[trim = {0cm 0cm 0cm 0cm}, clip, width=\textwidth]{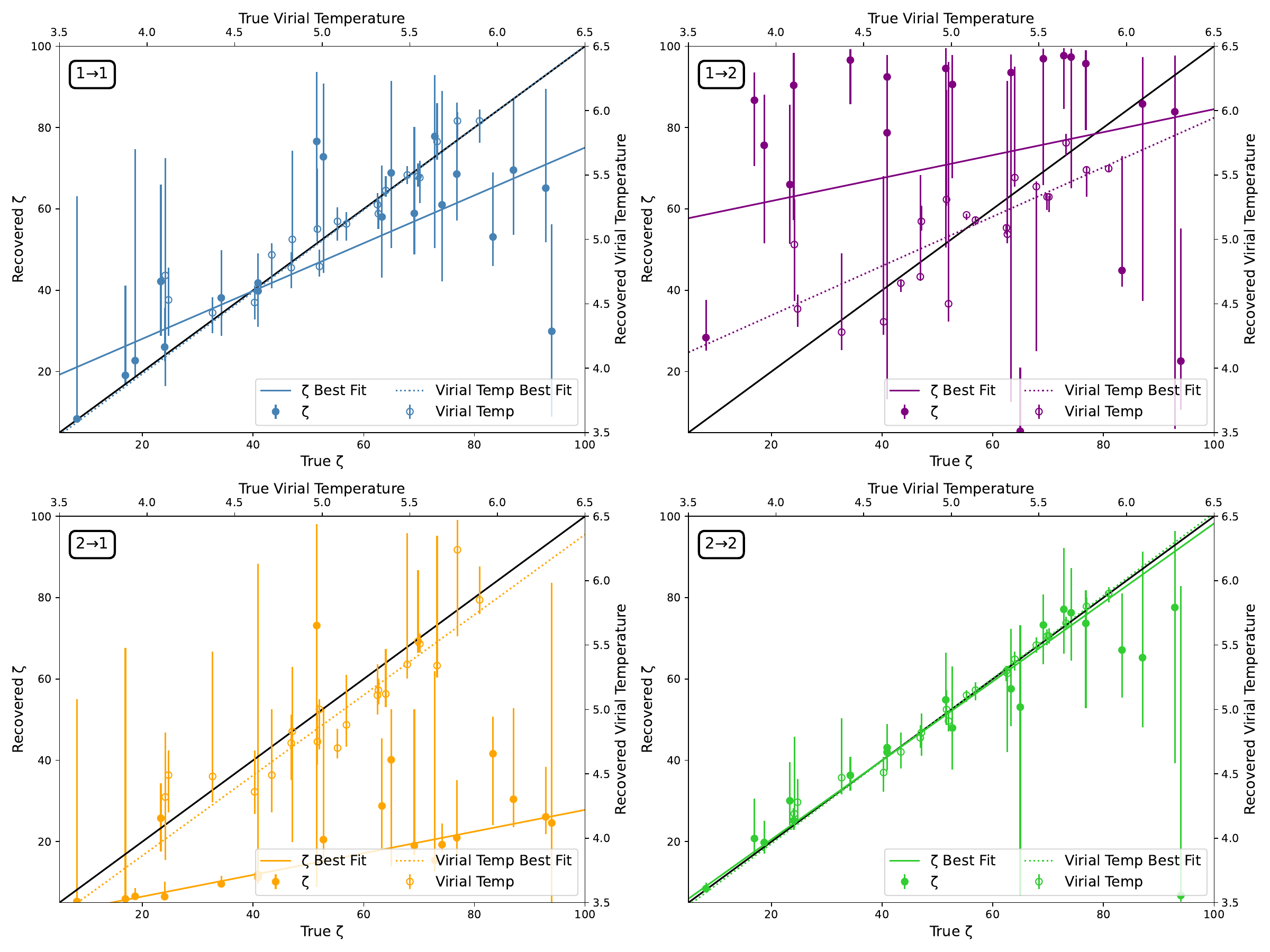}
\caption{Returned $\zeta$ and $T_{vir}^{min}$ values with the addition of 21cmSense noise for conditions 1$\rightarrow$1 (top left), 2$\rightarrow$2 (bottom right), 1$\rightarrow$2 (top right), and 2$\rightarrow$1 (bottom left). $\zeta$ is plotted with filled circles along the bottom and left axes, while $T_{vir}^{min}$ is plotted with open circles along the top and right axes. The solid and dashed lines show the best fit trend lines for $\zeta$ and $T_{vir}^{min}$, respectively.  The black line in each subplot represents a perfect recovery of $\zeta$ or $T_{vir}^{min}$ (i.e. y=x).}
\label{fig:noise}
\end{figure*}

In this section, we introduce noise to two datasets of twenty light-cones in order to determine if 21CMMC's model dependence is a product of inputting noise-free data. If 21CMMC's model dependence is a result of small differences between the PS of the two bubble-finding algorithms that fall below instrumental noise, then introducing noise to our input data may soothe the observed disagreement. We simulate noise using the package 21cmSense \citep{Pober13,Pober14,Murray2024}.\footnote{https://github.com/rasg-affiliates/21cmSense} 21cmSense outputs the sensitivities of a 21 cm PS as a function of $k$, meaning the noise is not added as a specific realization to each lightcone, but rather as an uncertainty included in the MCMC fits to the PS. Here we used the ``moderate'' foreground excision model, which uses a defined horizon buffer. We chose the moderate foreground excision model for this follow-up test as a more conservative representation of the current capabilities of radio interferometers than the ``optimistic'' foreground excision model, although a similar analysis performed with the optimistic model yielded comparable results. We further assume a HERA-like interferometer and the standard 21cmSense observation time of 1080 hours (180 observation days with 6 observing hours per day). Figure \ref{fig:noise_ps} shows our assumed uncertainties on each $k$-mode of the same bubble-finding algorithm 2 light-cone as in Figure \ref{fig:lightcones}. Points for whom 21cmSense outputted an infinite uncertainty were excised from our PS files. 

Figure \ref{fig:noise} shows both 21CMMC's returned $\zeta$ and $T_{vir}^{min}$ values with the addition of noise (because we have only 20 light-cones, and not 100, we plot both sets of parameters on the same axes). While all the fits are poorer than the noise-free case, we can see that 21CMMC's performance in returning $\zeta$ for conditions 1$\rightarrow$2 and 2$\rightarrow$1 (top right and bottom left), decreases significantly more than the degree to which its performance on conditions 1$\rightarrow$1 and 2$\rightarrow$2 decreased (top left and bottom right). No correlation is obvious between true and returned $\zeta$ in condition 1$\rightarrow$2, although the points continue to pile at the ceiling. A correlation appears to exist between true and returned $\zeta$ in condition 2$\rightarrow$1, but it is much weaker than its analog in Figure \ref{fig:main-zeta}. 

In Table \ref{tab:tab_3}, we see that the $\chi_{\nu}^{2}$ values universally decreased between Figures \ref{fig:main-zeta} and \ref{fig:noise}, but this is a product of the much larger uncertainties placed on each point and not on increased accuracy. Also worth noting is that condition 1$\rightarrow$2 again benefited from the ceiling effect.

Figure \ref{fig:noise} also shows 21CMMC's returned $T_{vir}^{min}$ values with the addition of noise, the analog of Figure \ref{fig:main-tvir}. This result well-mirrors Figure \ref{fig:main-tvir}--conditions 1$\rightarrow$1 and 2$\rightarrow$2 closely follow the identity line, while condition 1$\rightarrow$2 appears to be more randomly distributed about the identity line, likely an artifact of the ceiling effect discussed above. Condition 2$\rightarrow$1 again follows the identity line, but is outperformed by condition 1$\rightarrow$1. $\chi_{\nu}^{2}$ values again decreased as a result of significantly increased uncertainties on each point.

This result shows that 21CMMC's model dependence does not fall below instrumental noise, and in fact is exacerbated by the presence of noise.  This is largely due to the steep redshift scaling of the noise, which is driven by the spectral index of Galactic synchrotron emission. As discussed in Section \ref{sec:discussion}, the agreement between the two bubble finding algorithms is best at the highest redshifts (largely because there are no bubbles prior to the onset of reionization).  With the very large noise at high redshifts, the fits are instead driven by the lowest redshifts, exacerbating the model dependence.  As a simple test, we ran two analyses without noise but excluding redshifts below $z \approx 7.8$ and redshifts above $z \approx 11.5$, corresponding to the three lowest-redshift chunks and the three highest-redshift chunks of the light-cone respectively. The out-of-domain performance declined when the high redshift slices were excised, supporting the idea that the decline in performance when introducing instrumental noise is driven by a down-weighting of high redshifts.

\setlength{\tabcolsep}{18pt}
\begin{deluxetable*}{rllll}
\digitalasset
\tablecaption{$\zeta$ and $T_{vir}^{min}$ $\chi_{\nu}^{2}$ Values With 21cmSense Noise \label{tab:description}}
\tablehead{
\colhead{} & \colhead{1$\rightarrow$1} & \colhead{2$\rightarrow$2} & \colhead{1$\rightarrow$2} & \colhead{2$\rightarrow$1}
}
\startdata
$\zeta$ & 0.950 & 0.430 & 7.882 & 21.861 \\
\hline
$T_{vir}^{min}$ & 0.630 & 0.328 & 64.368 & 0.825 \\
\enddata
\tablecomments{$\chi^{2}_{\nu}$ values for $\zeta$ (top) and $T_{vir}^{min}$ (bottom) returns for conditions 1$\rightarrow$1, 2$\rightarrow$2, 1$\rightarrow$2, and 2$\rightarrow$1 with the addition of 21cmSense noise.}
\end{deluxetable*}
\label{tab:tab_3}

\section{Discussion}
\label{sec:discussion}

In this section, we present explanations for the shape of the power spectra in Figures \ref{fig:main-zeta} - \ref{fig:noise} and discuss more generally 21CMMC's performance on in-domain versus out-of-domain runs. In Section \ref{sec:degeneracy-explanation}, we discuss why the difference between bubble-finding algorithms 1 and 2 is degenerate with $\zeta$, and in Section \ref{sec:param_recovery}, we discuss 21CMMC's recovery of $\zeta$ and $T_{vir}^{min}$.

\subsection{Degeneracy Between $\zeta$ and Bubble-finding Algorithms}
\label{sec:degeneracy-explanation}

\begin{figure*}[ht!]
\includegraphics[scale=0.65, trim = {0cm 2cm 0cm 3cm}, clip, width=\textwidth]{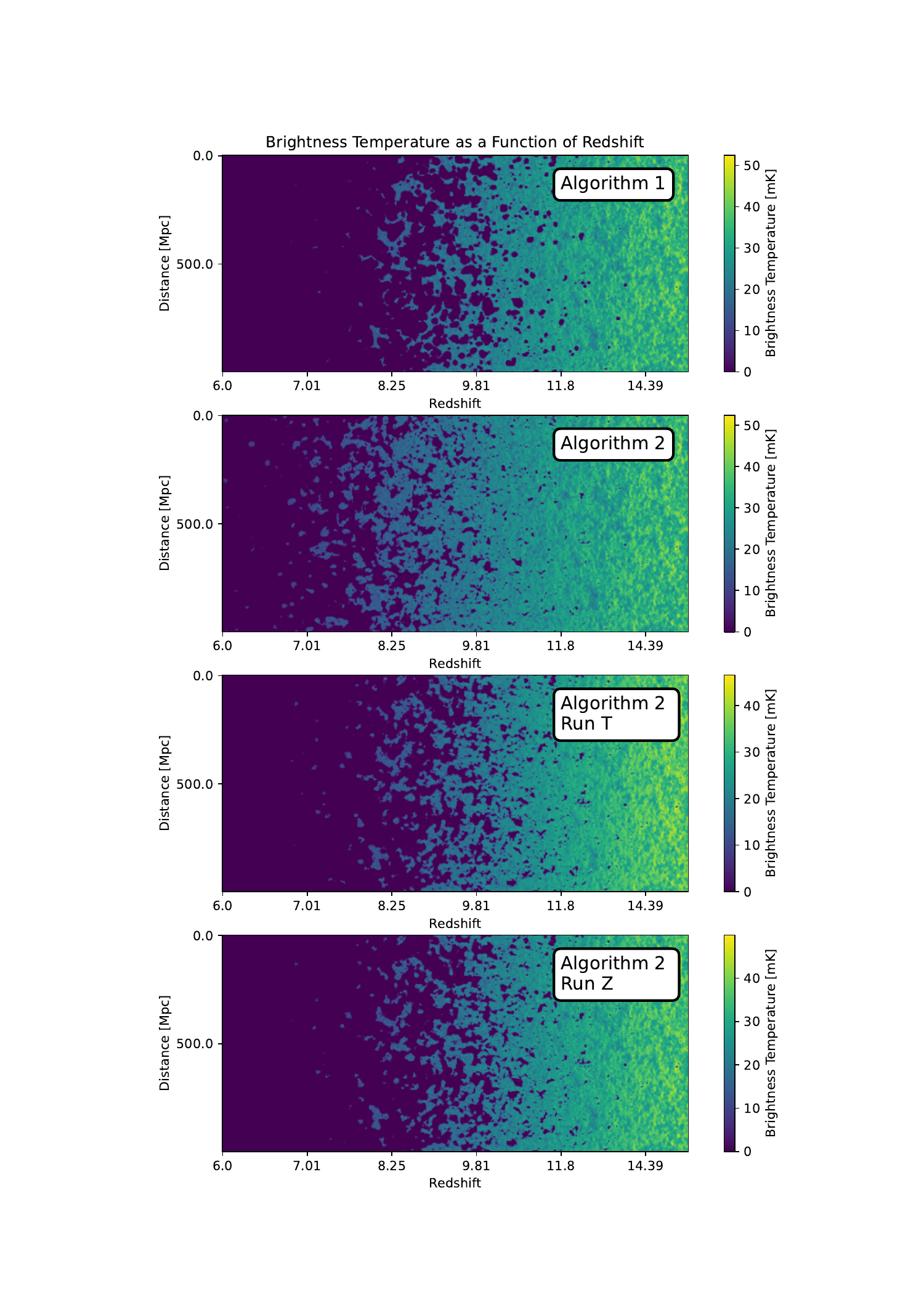}
\caption{Slices along one spatial axis of the brightness temperature maps for light-cones simulated using bubble-finding algorithm 1 (top), bubble-finding algorithm 2 (second from the top), bubble-finding algorithm 2 with a $T_{vir}^{min}$ value recovered from 21CMMC (second from the bottom), and bubble-finding algorithm 2 with a $\zeta$ value recovered from 21CMMC (bottom). The slices labeled ``Run T'' and ``Run Z'' correspond to the light-cones described as such in Section \ref{sec:PS_recreate}. The underlying density field was kept the same between the four light-cones.}
\label{fig:zeta-degeneracy}
\end{figure*}

It is worth addressing why the difference between bubble-finding algorithms 1 and 2 is degenerate with $\zeta$ in order to contextualize this work in the case of sampling the true 21 cm PS. Both $\zeta$ and $T_{vir}^{min}$ influence the speed of reionization, so it is surprising that the model dependence in returning the two parameters is so different in 21CMMC. One explanation for this fact comes from how the two parameters influence the speed of reionization. $T_{vir}^{min}$ is anticorrelated with the number of ionizing sources in a light-cone. In the context of a brightness temperature map, changing $T_{vir}^{min}$ would change the number of ionized bubbles in the map. This is clearly visible in Figure \ref{fig:zeta-degeneracy}. Figure \ref{fig:zeta-degeneracy} shows slices along one spatial axis of the brightness temperature maps for the light-cones generated in Runs T and Z as described in Section \ref{sec:PS_recreate}. These two slices are compared against the bubble-finding algorithm 1 light-cone used in Section \ref{sec:PS_recreate} as well as its bubble-finding algorithm 2 counterpart. The slice corresponding to Run T (second from the bottom) clearly has more ionized bubbles than the bubble-finding algorithm 2 control (second from the top). Meanwhile, $\zeta$ is correlated with the size of the ionized bubbles. This is also visible in Figure \ref{fig:zeta-degeneracy}, where the slice corresponding to Run Z (bottom) has larger bubbles than the bubble-finding algorithm 2 control. From this and our discussion of the difference between bubble-finding algorithms 1 and 2 (Section \ref{sec:algorithm_differences}), we can explain why $\zeta$ is degenerate with the difference between the bubble-finding algorithms. In broad strokes, bubble-finding algorithm 1 ionizes the same central pixels as bubble-finding algorithm 2 with all else equal, meaning the two algorithms ionize the same number of bubbles. Bubble-finding algorithm 1, however, ionizes more pixels about each central pixel, leading to larger ionized bubbles. Thus 21CMMC recreates the topology of the EoR in out-of-domain runs by adjusting $\zeta$ and not $T_{vir}^{min}$. This, too, is clearly visible in Figure \ref{fig:zeta-degeneracy}, where the slice corresponding to Run Z best reproduces the topology of the slice corresponding to bubble-finding algorithm 1 (top).

It is important not to obscure the bigger picture in our discussion of $\zeta$'s degeneracy. This specific algorithmic disagreement can be modeled by adjusting $\zeta$, but in all likelihood, other algorithmic disagreements exist which are degenerate with $T_{vir}^{min}$ or some other astrophysical parameter. More importantly, we do not know which parameters or combination of parameters are degenerate with the difference between the bubble-finding algorithms and the true 21 cm PS. The decline in performance between conditions 1$\rightarrow$1 and 2$\rightarrow$1 and conditions 2$\rightarrow$2 and 1$\rightarrow$2 for both $\zeta$ and $T_{vir}^{min}$ highlights an issue in using 21CMMC to constrain the astrophysics of the EoR. Using 21CMMC to constrain the astrophysics of the EoR without taking its dependence on its specific sampler into account neglects a potentially significant source of uncertainty and bias. 

\begin{figure*}[ht!]
\includegraphics[trim = {0cm 0cm 1cm 0cm}, clip, width=\textwidth]{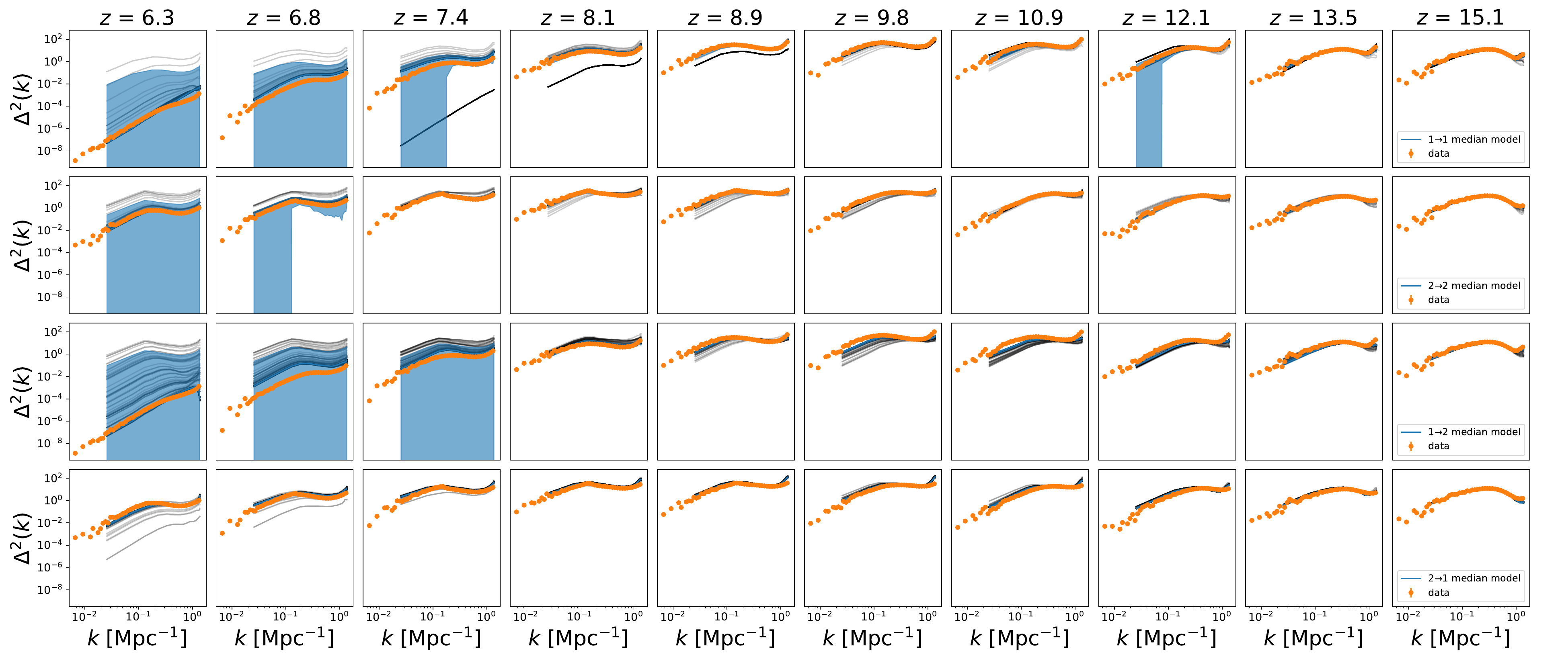}
\caption{Comparison of input PS and 21CMMC best-guess PS for conditions 1$\rightarrow$1 (top), 2$\rightarrow$2 (second from the top), 1$\rightarrow$2 (second from the bottom), and 2$\rightarrow$1 (bottom). The input PS is plotted in orange, each model is plotted in gray, and the median model at each chunk is plotted in blue. The blue shaded regions correspond to the standard deviation of the models at each $k$-mode. The light-cone pair used in this analysis is the same as in Figures \ref{fig:lightcones} and \ref{fig:ps-compare}.}
\label{fig:condition_comp}
\end{figure*}

Important to note is that, in nearly all cases, 21CMMC does a good job fitting the input power spectrum. Figure \ref{fig:condition_comp} shows a comparison of 21CMMC's best-fit PS with the input PS for conditions 1$\rightarrow$1 (top), 2$\rightarrow$2 (second from the top), 1$\rightarrow$2 (second from the bottom), and 2$\rightarrow$1 (bottom). The light-cone pair used in Figure \ref{fig:condition_comp} is the same as in Figures \ref{fig:lightcones} and \ref{fig:ps-compare}. Visually, the out-of-domain runs perform only marginally worse than the in-domain runs in fitting the input PS, and yet the out-of-domain runs perform much worse than the in-domain runs in returning $\zeta$ and $T_{vir}^{min}$. This shows that 21CMMC may fit the shape of the PS while simultaneously outputting parameters that are inaccurate to the input data.

21CMMC's modeling uncertainty warrants a discussion here. As described in \citet{GreigMes15}, 21CMMC places a ``modeling uncertainty'' on the input 21 cm PS proportional to $\Delta^2(k)$. This uncertainty is introduced to account for ``differences in the implementation of Radiative Transfer (RT) and errors in seminumerical approximations.'' 21CMMC's modeling uncertainty was introduced in order to account for precisely the problem outlined in this work, and the specific proportionality of 25$\%$ was chosen because a 21cmFAST PS differed from that of an RT simulation by 10-30 percent in \citet{zahn11}. However, as made clear by this work, a 10-30 percent difference in the PS can manifest itself as a much larger difference in 21CMMC's returned parameters, and the modeling uncertainty is insufficient in addressing 21CMMC's model dependence. Note that \citet{GreigMes15} used a modeling uncertainty of 25$\%$, whereas we used a modeling uncertainty of 15$\%$ in this work in agreement with the default value in 21CMMC at time of writing. We explored the consequences of adopting a 25$\%$ modeling uncertainty, and our general results were unchanged.

\subsection{Parameter Recovery}
\label{sec:param_recovery}

In this section, we discuss our recovered $\zeta$ and $T_{vir}^{min}$ values in our main result and follow-up tests. 

That condition 1$\rightarrow$2 categorically overestimates $\zeta$ in Figures \ref{fig:main-zeta} and \ref{fig:noise} is predictable given our discussions in Sections \ref{sec:algorithm_differences} and \ref{sec:degeneracy-explanation}. Reionization proceeds quicker in bubble-finding algorithm 1 light-cones than in bubble-finding algorithm 2 light-cones, so given $\zeta$'s proportionality with the speed of reionization and its degeneracy with the difference between the bubble-finding algorithms, 21CMMC prescribes an inflated value of $\zeta$ to a bubble-finding algorithm 1 light-cone when sampling bubble-finding algorithm 2. For a bubble-finding algorithm 2 light-cone to ionize as fast as a bubble-finding algorithm 1 light-cone, all else being equal, it must have a higher value of $\zeta$. The underestimation of $\zeta$ in condition 2$\rightarrow$1 follows from this observation. For a bubble-finding algorithm 1 light-cone to ionize as slowly as a bubble-finding algorithm 2 light-cone, all else being equal, it must have a lower value of $\zeta$.

Condition 2$\rightarrow$1 in Figures \ref{fig:main-tvir} and \ref{fig:noise} supports our hypothesis surrounding the different roles of $T_{vir}^{min}$ and $\zeta$ in the topology of reionization. Condition 2$\rightarrow$1 closely follows the identity line in Figures \ref{fig:main-tvir} and \ref{fig:noise}. As mentioned above, this is in line with our expectations--the difference between the two bubble-finding algorithms is degenerate with $\zeta$. It is again worth mentioning, however, that condition 1$\rightarrow$1 outperforms condition 2$\rightarrow$1 in both Figures \ref{fig:main-tvir} and \ref{fig:noise}, suggesting that even in returning $T_{vir}^{min}$, 21CMMC exhibits model dependence. This could be a result of subtle differences in small-scale structure between algorithm 1 and 2 light-cones. 

The results for condition 1$\rightarrow$2 in Figures \ref{fig:main-tvir} and \ref{fig:noise} are surprising given the performance of condition 2$\rightarrow$1 discussed above. One explanation for 21CMMC's poor performance in returning $T_{vir}^{min}$ in this case is that it is a consequence of the ceiling effect visible in condition 1$\rightarrow$2's $\zeta$ returns (see Figures \ref{fig:main-zeta} and \ref{fig:noise}). As the condition 1$\rightarrow$2 walkers traversed parameter space, it is possible that by imposing a ceiling of $\zeta = 100$, the walkers were forced to settle into a false minima, resulting in inaccurate returns of $T_{vir}^{min}$. Had the condition 1$\rightarrow$2 walkers been given a broader slice of parameter space to explore, allowing them to settle into their preferred $\zeta$ value away from the ceiling, they may have been better able to constrain $T_{vir}^{min}$. 

The general trend of 21CMMC's recovery of $\zeta$ is that out-of-domain runs do not follow the identity line, while in-domain runs do. 21CMMC's model dependence is clearest in the recovered values of $\zeta$, and the results in Figure \ref{fig:main-zeta}, in conjunction with those results of the follow-up-tests and our discussion in Section \ref{sec:degeneracy-explanation} suggest that 21CMMC is unable to accurately constrain $\zeta$ in out-of-domain runs. The trend of 21CMMC's out-of-domain $T_{vir}^{min}$ recovery, meanwhile, is that it exhibits a weak model dependence. This dependence would likely be strengthened were we to sample $T_{vir}^{min}$ alongside some other astrophysical parameter, as demonstrated by Table \ref{tab:tab_2}. If, for instance, we had sampled $T_{vir}^{min}$ alongside a parameter uncorrelated with the speed of reionization, 21CMMC would have exhibited a strong model dependence in returning $T_{vir}^{min}$.

Were 21CMMC's model dependence limited to returning 21cmFAST specific parameters like $\zeta$, this issue would be limited in scope. However, Section \ref{sec:observables} reveals that 21CMMC's model dependence impacts its ability to return the in-principle model-independent parameters $\Delta z$ and $z_{50}$, the duration and midpoint of reionization respectively. Although the analysis presented in Section \ref{sec:observables} did not fully propagate errors, 21CMMC demonstrated a clear bias in returning duration of reionization in out-of-domain runs. Such a result calls 21CMMC's use case into question. 

\section{Conclusion}
\label{sec:conclusion}

21CMMC is a publicly available MCMC sampler of 21cmFAST, a seminumerical simulator of the 21 cm signal. Among other free parameters, 21cmFAST allows the user to pick from two ``bubble-finding'' algorithms in the simulation, described in \citet{Mesinger07}, \citet{Zahn07}, \citet{zahn11}, and \citet{Mesinger11}. The two algorithms differ in their treatment of ionized regions of the IGM, with bubble-finding algorithm 1 flagging the entire region satisfying the ionization criteria while bubble-finding algorithm 2 ionizes only the central pixel. This difference manifests itself as a different rate of reionization, among other topological differences, between the two algorithms with bubble-finding algorithm 1 light-cones ionizing quicker than bubble-finding algorithm 2 light-cones with all else being equal. Which algorithm is the more physically accurate is unclear, however, algorithm 2 is the default due to its computational efficiency.

In this work, we examined 21CMMC's model dependence between the sampler and the input data. We do so by performing in-domain and out-of-domain runs of 21CMMC on 100 pairs of light-cones, which differ only by whether they were simulated with bubble-finding algorithm 1 or 2. We then compare recovered values of $\zeta$ and $T_{vir}^{min}$ between runs. We find that 21CMMC is unable to accurately return $\zeta$ for out-of-domain runs. We also find that 21CMMC suffers a weak model dependence in recovering $T_{vir}^{min}$. That 21CMMC returns $T_{vir}^{min}$ somewhat accurately is likely a consequence of our choice of parameterization and not a quality of $T_{vir}^{min}$. Meanwhile the strong dependence 21CMMC exhibits in returning $\zeta$ is due to the fact that the precise difference between the two bubble-finding algorithms is degenerate with $\zeta$. This results in 21CMMC being completely unable to return $\zeta$ in out-of-domain runs. Had we chosen not to sample $\zeta$, however, and varied $T_{vir}^{min}$ alongside a parameter uncorrelated with the speed of reionization, 21CMMC likely would have exhibited a stronger model dependence in returning $T_{vir}^{min}$ than shown in this work.

We then assessed 21CMMC's ability to return the in-principle model-independent duration and midpoint of reionization. We found that although 21CMMC was able to accurately return the midpoint of reionization, it demonstrated a clear bias in returning the duration of reionization. This result suggests that the impact of 21CMMC's model dependence is greater than previously expected. Work must be done towards understanding 21CMMC's model dependence before it can be trusted in constraining any astrophysical parameters of the EoR, specific to 21cmFAST or otherwise.

We next ran 21CMMC with the addition of realistic noise calculated for HERA using 21cmSense to determine if 21CMMC's model dependence falls below instrumental noise. While the addition of noise degrades all constraints, we observe that 21CMMC's performance in out-of-domain runs declines more than it declines for in-domain runs, demonstrating that 21CMMC's model dependence will likely be a problem for the current generation of radio interferometers, and not only extremely sensitive future interferometers.

While we have focused on the bubble finding algorithm, this work has several important consequences that suggest further exploration of seminumerical 21 cm algorithms is necessary.  Even with the bubble finding algorithm removed in the recent 21cmFAST version 4 update, there are many other modeling assumptions implicit in the approach.  Multiple other algorithms exist beside 21cmFAST, e.g., SIMFAST21 \citep{santos_et_al_2010}, zreion \citep{battaglia_et_al_2013}, AMBER \citep{trac_et_al_2022}, GRIZZLY \citep{ghara_2023}, etc.  These are all likely to have algorithmic differences akin to (or more significant than) the two bubble finding algorithms explored here.  \citet{zhou_and_la_plante_2022} have already shown that Machine Learning models trained on 21cmFAST simulations fail to recover the CMB optical depth from reionization when given zreion simulations to analyze (and vice versa). \citet{jasper_paper} showed that AI models trained on a single seminumerical model fail to recover the duration of reionization of a second unseen model, albeit out-of-distribution performance improves as the training data is diversified. Similar studies are difficult to perform with 21CMMC since, for example, zreion does not have ``true'' values of $\zeta$ and $T_{vir}^{min}$ to recover (as these parameters are unique to 21cmFAST) --- hence the value of the bubble finding algorithms within 21cmFAST for this initial study.

Some of these algorithms are almost certainly more correct than others, but in many ways that is besides the point.  Ultimately, the real Universe will exhibit more complicated physics than can be captured by the parameterization of any seminumerical algorithm.  While our present analysis is limited in scope, it illustrates the work to be done to ensure that MCMC analyses using seminumerical models can be used to robustly interpret upcoming 21 cm data without concern for hidden algorithmic dependencies.

\begin{acknowledgements}
The authors acknowledge support from NSF award 2106510 and the associated Research Experiences for Non-Traditional Undergraduates (RENTU) program and the NASA Rhode Island Space Grant Program, award 80NSSC25M7082. Part of this research was conducted using computational resources and services at the Center for Computation and Visualization, Brown University. A.B. also thanks Willow Smith for mentorship during the nascent stages of this project and Jasper Solt for debugging code throughout the project.  The authors also thank the anonymous referee for helpful suggestions that improved the quality of this manuscript.  
\end{acknowledgements}

\bibliography{bibliography}{}
\bibliographystyle{aasjournal}

\end{document}